\documentclass[a4paper,14pt]{extarticle}
\usepackage{cmap} 
\usepackage[cp1251]{inputenc}
\usepackage[english,russian]{babel}

\usepackage{braket,amssymb,amsmath,array,commath,mathtext,wrapfig,tikz}
\usepackage[arrows=pgf-filled,version=3]{mhchem}
\usepackage{bm}
\usepackage{amsthm,amsfonts,amscd}
\usepackage{graphicx}
\usepackage{cite}
\usepackage{lscape}
\usepackage{geometry}
\usepackage{indentfirst}
\usepackage{afterpage,soul,ulem,dcolumn,changebar,longtable,hhline,multirow,makecell}
\usepackage[small,bf]{caption}
\makeatletter
\renewcommand{\@biblabel}[1]{#1.\hfil} 
\makeatother

\begin{document}
\renewcommand{\refname}{References}
\textit{The Forces outside the static limit in the rotating frame} \\
\textbf{Andrei Grib$^{1}$\,, Vitalii Vertogradov$^{1,2}$\,, Ivan Fedorov$^{1}$}\\

$^1$ Herzen State Pedagogical University of Russia, 48 Moika Emb., Saint Petersburg 191186, Russia \\
$^2$ SPb branch of SAO RAS, 65 Pulkovskoe Rd, Saint Petersburg 196140, Russia \\
Andrey Grib andrei\_grib@mail.ru \\
 Vitalii Vertogradov vdvertogradov@gmail.com \\

\textbf{Abstract}
In this paper we consider the rotating frame of the Minkowski spacetime in order to describe the inertial forces outside the static limit. We consider the inertial forces inside the static limit to find the classical analogue afterwards we find out the expressions for these forces outside the static limit where we can't consider the limit $\frac{v}{c}\to 0$. We show that in general case if the angular velocity of the object $\Omega$ is equal to the angular momentum $\omega$ then the acceleration is equal to zero.

\section*{Introduction}

Consideration of the rotating black holes led the authors of~\cite{bib:grib} to conclusion that such notions as the ergosphere, existence of the geodesics with negative energies, Penrose effect~\cite{bib:pen, bib:ver, bib:pavlov} are features typical not only for black holes but also exist even in Minkowski spacetime in rotating coordinates. It was shown that the role of the static limit in the theory of  rotating  black holes plays the distance from the origin of rotating coordinates to the value of the radius where the linear velocity of rotation becomes equal to the speed of light. The region out of this value plays the role of ergosphere. In this region all bodies must be moving.This corresponds to use of the rotating rods and this is the main difference with observations inside the limit where all bodies can be at rest.
In this paper we consider the geodesic lines as inside the static limit in rotating coordinates as outside of it. This is necessary for discussion of the inertial forces appearing in these coordinates in Minkowski spacetime. These forces inside the limit are well known centrifugal and Coriolis forces. The analogues of these forces for the region outside of the static limit are found in this paper.

In the paper~\cite{bib:grib} authors showed that the analogue of the Penrose effect outside the static limit can lead to observational effects inside this static limit. The main goal of the current paper is the investigation of a new phenomenon - the inertial forces outside the static limit and their difference with those inside the static limit. It is necessary to take into account the relativistic corrections  when one considers the movement of the spacecrafts in this region, by using the rotating frame and observing the signals from them on the Earth.

In this paper Latin indices take values $0, 1, 2, 3$ and Greek indices take values $1, 2, 3$.

\section{Inside the static limit}
\setcounter{equation}{0}

In this paper we consider the flat Minkowski spacetime in cylindrical coordinates  $\{ ct\,, r\,, \varphi' \,, z \}$:
\begin{equation}
ds^2=c^2dt^2-dr^2-r^2d\varphi'^2-dz^2 \,.
\end{equation}

 If we consider the rotating coordinates in Minkowski spacetime 
\begin{equation}
\varphi'=\varphi - \omega t \,,
\end{equation}
then we obtain the following line element:

\begin{equation}\label{eq:metric}
ds^2=c^2dt^2-r^2(d\varphi -\omega dt)^2-dr^2-dz^2 \,.
\end{equation}

Here $\omega$ is angular velocity. 

One can note that the metric component $g_{00}$ changes its sign at $r=r_{sl}=\frac{c}{\omega}$. However, the line element \eqref{eq:metric} might still be timelike due to the positivity of the off-diagonal term $g_{02}dtd\varphi=\omega r^2dtd\varphi$. If we compare the metric \eqref{eq:metric} with the Kerr metric then the hypersurface $r=r_{sl}$ plays a role of the static limit in Kerr black hole. 

The Killing vector outside the static limit has different form than inside the static limit i.e. $\frac{d}{dt}+\omega \frac{d}{d\varphi}$. Due to this fact we have another expression for the proper time where off-diagonal term is not equal to zero. Thus, the proper time will be imaginary for a body at rest. In this region and the geodesic equations will lead to new phenomenon - the inertial forces.

If we consider the observer on the Earth (which is motionless and everything is rotating around it) then the object which is situated at $r>r_{sl}$ can't be static for our observer. It must also rotate otherwise the line element \eqref{eq:metric} will be spacelike. Also in this case the proper time will be imaginary because if we consider the connection between the proper $\tau$ and coordinate $t$ times when $r=const. \,, \varphi =const. \,, z=const.$ then we obtain:

\begin{equation}\label{eq:owntime}
d\tau=\sqrt{g_{00}}dt \,,
\end{equation}
and we know that $g_{00}<0$ at $r>r_{sl}$.

So if we want to calculate inertial forces for the objects inside the static limit using \eqref{eq:owntime} we have to consider only the region $ 0\leq r \leq r_{sl}$. In this region $g_{00}>0$. From the Newtonian mechanics  we know that an acceleration is proportional to a force. So to calculate forces one should use the right hand-side of the geodesic equation:
\begin{equation}\label{eq:cris}
\frac{du^i}{d\tau}=-\Gamma^i_{kl}u^ku^l \,,
\end{equation}
where $\Gamma^i_{kl}$ is the Cristoffel symbol.

One should note that an acceleration is the time derivative of the 4-velocity $u^i$. From the the differential geometry course we know that the derivative doesn't obey the tensor transformation law. So when we define a force we should use the covariant derivative instead of usual one. We want to calculate the three-force so we should use the three covariant derivative instead of \eqref{eq:cris}. In our case we have~\cite{bib:landau}:
\begin{equation}
f^\alpha=u^\alpha_{;\beta}u^\beta=\frac{du^\alpha}{d\tau}+\gamma^\alpha_{\beta \delta}u^\beta u^\delta \,,
\end{equation}
where $\gamma^\alpha_{\beta \delta}$ is the purely three-Cristoffel symbol. So we can note (and show below) that the terms in the right hand-side \eqref{eq:cris} which are proportional to $u^0$ correspond to the centrifugal force and those which are proportional to $u^0u^\alpha$ correspond to the Coriolis force and those which are proportional to $u^\alpha u^\beta$ are the part of three covariant derivative and are not forces at all. 

We should know Cristoffel symbol of the metric \eqref{eq:metric} to calculate the geodesic equation \eqref{eq:cris}. Using the well-know formula:

\begin{equation}
\Gamma^i_{kl}=\frac{1}{2}g^{ij}\left( g_{kj,l}+g_{jl,k}-g_{kl,j}\right) \,,
\end{equation}
one can obtain non-vanishing component of Cristoffel symbols:
\begin{equation}
\begin{split}
\Gamma^1_{00}=-\frac{\omega^2 r}{c^2} \,, \\
\Gamma^1_{02}=\frac{\omega r}{c}  \,, \\
\Gamma^1_{22}=-r \,, \\
\Gamma^2_{12}=\frac{1}{r} \,, \\
\Gamma^2_{01}=-\frac{\omega}{cr} \,.
\end{split}
\end{equation}

 To calculate the inertial forces in the region $0\leq r\leq r_{sl}$ we can use the well-known formula~\cite{bib:landau}:

\begin{equation} \label{eq:f}
F_\alpha =\frac{mc^2}{\sqrt{1-\frac{v^2}{c^2}}}\left [ -\frac{d}{dx^\alpha} \left (\ln \sqrt{\left( 1-\frac{\omega^2 r^2}{c^2} \right )} \right ) +\sqrt{1-\frac{\omega^2 r^2}{c^2}}\left ( \frac{d\xi_\beta}{dx^\alpha}-\frac{d\xi_\alpha}{dx^\beta}\right ) \frac{v^\beta}{c} \right ] \,,
\end{equation}
where
\begin{equation} \label{eq:xi}
\begin{split}
\xi_\alpha=\frac{g_{0\alpha}}{1-\frac{\omega^2 r^2}{c^2}} \,, \\
v^\alpha= \frac{cdx^\alpha}{\sqrt{1-\frac{\omega^2 r^2}{c^2}}\left( cdt-\xi_\beta dx^\beta\right) } \,.
\end{split}
\end{equation}

Note that the first term in the square brackets corresponds to the centrifugal force and the second term depends on the velocity linearly and corresponds to the Coriolis force. 

Substituting \eqref{eq:xi} into \eqref{eq:f} we obtain:

\begin{equation} \label{eq:radial_forse}
F_r=\left[ \frac{mc^2}{\sqrt{1-\frac{v^2}{c^2}}} \left( \frac{2\omega^2 r}{c^2-\omega^2 r^2} \right )\right ]_{centr}+\left[ \frac{2\omega r }{\left ( 1-\frac{\omega^2 r^2}{c^2}\right )^2 } \frac{d\varphi}{cdt-\xi_2d\varphi} \right]_{cor} \,,
\end{equation}

\begin{equation} \label{eq:azimutal}
F_\varphi= \left[ \frac{2\omega r }{\left ( 1-\frac{\omega^2 r^2}{c^2}\right )^2 } \frac{dr}{cdt-\xi_2d\varphi} \right]_{cor} \,.
\end{equation}

Let's find out where the static limit $r_{sl}$ is situated in the Solar system. For this purpose let's consider two cases:

\begin{enumerate}
\item The static observer is on the surface of the Earth and everything is rotating around it. 
In this case we should solve the following algebraic equation:
\begin{equation} \label{eq:omega}
1-\frac{\omega^2r^2}{c^2}=g_{00}=0 \,.
\end{equation}
In this case the period is equal to 24 h and the static limit is situated between the Uranus and the Neptune.
\item Everything is rotating around the Sun and the observer is on the Earth. 
In this case the period is equal to 1 year and $r_{sl}=1.5*10^{15} \, m$. This static limit is situated inside the Oort's clouds. 
\end{enumerate}

\section{General case}
\setcounter{equation}{0}

From now and on we consider the system in units $c=1$.

In general case the proper time $\tau$ has the following form:

\begin{equation} \label{eq:proper_time}
d\tau^2=(1-\omega^2r^2)dt^2+2\omega r^2dtd\varphi-dr^2-dz^2-r^2d\varphi^2 \,.
\end{equation}

One can see that the square of proper time \eqref{eq:proper_time} might be positive in the region $r_{sl}=\frac{1}{\omega}<r<+\infty$ because of the off-diagonal term $+2\omega r^2 dtd\varphi$. Hence we state that we can consider the region $r_{sl}<r<+\infty$ only in the case when the observer has non-vanishing angular velocity $\Omega=\frac{d\varphi}{dt}$. However, we assume that $r= const.$ and $z=const.$ for the observer and the connection between the proper and coordinate times is the following:

\begin{equation} \label{eq:connection}
d\tau=dt\sqrt{1-r^2(\Omega-\omega)^2} \,.
\end{equation}

To obtain force expression we should write down geodesic equations which are given by in general case:
\begin{equation}
\begin{split} \label{eq: geodesic}
\frac{d^2r}{d\tau^2}=\omega ^2r \left (\frac{dt}{d\tau} \right )^2-2\omega r \frac{dt}{d\tau}\frac{d\varphi}{d\tau}+r\left( \frac{d\varphi}{d\tau} \right )^2 \,, \\
\frac{d2\varphi}{d\tau^2}=-\frac{2}{r}\frac{dr}{d\tau}\frac{d\varphi}{d\tau}+2\frac{\omega}{r}\frac{dt}{d\tau}\frac{dr}{d\tau} \,,
\end{split}
\end{equation}

If we followed the force definition from the previous section then we would see that forces have finite values on the static limit but they are extremely large. No one sees these large values so we should redefine forces. 

In the previous section we defined forces as three covariant derivative of the momentum and didn't count some terms in the right-hand side of the equation \eqref{eq:geodesic} because they were the part of the three covariant derivative. However, in general case, outside the static limit, the Killing vector $\frac{d}{dt}$ is spacelike and as the result unphysical. In the region $r>r_{sl}$ the observer, like in the ergoregion of a rotating black hole, has to move along both $t$ and $\varphi$ coordinates.As the result of such movement all Cristoffel symbols proportiaonal to $(u^0)^2 \,, u^0u^\varphi$ and $(u^\varphi)^2$ are inertial forces. So substituting \eqref{eq:connection} into \eqref{eq:geodesic} one obtains:
\begin{equation} \label{eq:usk}
\begin{split}
\frac{d^2r}{d\tau^2}=\frac{r}{1-r^2(\omega-\Omega)^2} \left( \Omega-\omega \right)^2 \,, \\
\frac{d^2\varphi}{d\tau^2}=\frac{2v_r}{r-r^3(\omega-\Omega)^2}\left (\omega-\Omega \right ) \,, \\
v_r=\frac{dr}{dt} \,.
\end{split}
\end{equation}

From \eqref{eq:usk} one can see that the object has non-vanishing acceleration only if $\omega-\Omega \neq 0$. The object doesn't have any acceleration if $\Omega=\omega$ because in this case it is equal to zero. If $\omega \neq \Omega$ then the acceleration is groing function of $r$ and at the surface $r=\frac{1}{\omega-\Omega}$ diverges. This situation is the same like in previous section with the angular velocity $\omega$ is being replaced by $\omega-\Omega$. If the acceleration of an object is a constant $W$ then the $\Omega =\frac{d\varphi}{dt}$ is the decreasing function of $r$ i.e.:
\begin{equation}
\Omega=\omega-\sqrt{\frac{W}{r+Wr^2}} \,.
\end{equation}

So one can see that with the proper time \eqref{eq:connection} we should redefine inertial forces. It leads us to the fact that if $\Omega=\omega$ then the acceleration is absent. However, if the body has the constant acceleration then it means that $\Omega \neq \omega$ but $\omega-\Omega$ tends to zero with growing $r$. Also one should notice that $\Omega$ must be always positive otherwise all metric components become negative outside the static limit and the line element \eqref{eq:metric} would be spacelike in this region. However, in the region $0\leq r\leq r_{sl}$ we can consider negative values of $\omega$.

\section{Conclusion}
\setcounter{equation}{0}

In this paper we have considered forces inside and outside the static limit in Minkowski spacetime in the case of rotating coordinates. In this frame one has only two types of inertial forces i.e. centrifugal and the Coriolis forces. In the case of the static bodies one can consider forces only up to the radius $r=r_{sl}$ because in this case the proper time and velocity is imaginary outside the static limit. So to consider the inertial forces in the region $r_{sl}\leq r\leq \infty$ one should consider the general proper time with non-vanishing angular velocity $\Omega =\frac{d\varphi}{dt}$. We have found out that in the case of the proper time \eqref{eq:connection} the acceleration tends to zero if $\Omega \to \omega$. Also one should notice that if $\omega\neq\Omega$ then we have another surface $r=\frac{1}{\omega-\Omega}$. This surface shows that outside it the line element \eqref{eq:metric} is spacelike. Or, the case $\omega\neq\Omega$ is analogue of the static observer with the angular momentum of spacetime $\omega-\Omega$. It is worth mentioning that all forces which are considered in this article are fictitious ones.

\section{Acknowledgments} 
The work was performed within the SAO RAS state assignment in the part "Conducting Fundamental Science Research".

\end{document}